\documentclass[aip,jcp,preprint]{revtex4-2}

\usepackage{amsmath}
\usepackage{longtable}
\usepackage{booktabs}
\usepackage{graphicx}
\usepackage[section]{placeins}

\makeatother

\begin{document}

\title{Accurate and Efficient Prediction of Double Excitation Energies Using the Particle-Particle Random Phase Approximation}

\author{Jincheng Yu}
\thanks{These two authors contributed equally}
\affiliation{Department of Chemistry, Duke University, Durham, North Carolina 27708, United States}
\author{Jiachen Li}
\thanks{These two authors contributed equally}
\affiliation{Department of Chemistry, Yale University, New Haven, Connecticut 06520, United States}
\author{Tianyu Zhu}
\affiliation{Department of Chemistry, Yale University, New Haven, Connecticut 06520, United States}
\author{Weitao Yang}
\email{weitao.yang@duke.edu}
\affiliation{Department of Chemistry, Duke University, Durham, North Carolina 27708, United States}

\begin{abstract}
Double excitations are crucial to understanding numerous 
chemical, physical, and biological processes, 
but accurately predicting them remains a challenge. 
In this work, 
we explore the particle-particle random phase approximation (ppRPA) 
as an efficient and accurate approach for 
computing double excitation energies. 
We benchmark ppRPA using various exchange-correlation functionals 
for 21 molecular systems and two point defect systems. 
Our results show that ppRPA with functionals containing 
appropriate amounts of exact exchange
provides accuracy comparable to high-level wave function methods 
such as CCSDT and CASPT2, 
with significantly reduced computational cost. 
Furthermore, we demonstrate the use of ppRPA starting from an excited ($N-2$)-electron state calculated by $\Delta$SCF for the first time, as well as its application to double excitations in bulk periodic systems.
These findings suggest that ppRPA is a promising tool 
for the efficient calculation of 
double and partial double excitation energies in both molecular and bulk systems.
\end{abstract}

\maketitle

\section{INTRODUCTION}
%=INTRO-I SIGNIFICANCE OF DOUBLE EXCITATION============================
Double excitations,
where two electrons are promoted from occupied orbitals to 
virtual orbitals of the reference state,
play a significant role in understanding many important 
chemical, physical and biological processes,
including
the photochemistry of conjugated polymers\cite{
caveTheoreticalInvestigationSeveral1988,
caveDressedTDDFTTreatment2004,
lappeVerticalAdiabaticExcitation2000,
maitraDoubleExcitationsTimedependent2004,
starckeHowMuchDouble2006,
mazurDressedTDDFTStudy2011,
angeliAnalysisDynamicPolarization2010},
formations of the triplet-pair state in the singlet fission\cite{smithSingletFission2010}
and conical intersections in coupled electron-nuclear 
dynamics\cite{levine2006conical}.
%=EXPERIMENTAL METHODS=================================================
Experimentally,
directly studying double excitations with spectroscopic methods 
is challenging, 
because they are dark and usually cannot be observed in photo-absorption spectroscopy\cite{loosDynamicalCorrectionBethe2020}.
Therefore,
the optically forbidden doubly excited states can only be indirectly probed by methods
such as photoluminescence\cite{millerBiexcitonsGaAsQuantum1982}
and spectrally resolved four-wave mixing experiments\cite{
feuerbacherBiexcitonicContributionDegeneratefourwavemixing1991}.

%=CHALLENGES FOR THEORETICAL CALCULATIONS==============================
The theoretical study of double excitations involves
the quasiparticle state named biexciton that is composed of 
two particles and two holes.
The accurate description of biexcitons requires computing 
four-body correlations\cite{kossoskiReferenceEnergiesDouble2024a},
which is a crucial for describing simultaneous propagation 
of two electrons or two holes.
However, the calculation of four-body correlations is a difficult task in the theoretical treatment 
when many-body effects are important.
%Conventional single-reference methods face significant challenges
%when modelling double excitations.
This leads to significant challenges for single-reference methods to model double excitations.
%=====DFT-BASED METHODS================================================
For example,
the adiabatic time-dependent density functional theory (TD-DFT)\cite{
casida1995time,ullrich2011time},
which considers only singly excited states that consist of one particle and one hole,
cannot be applied for the computation of doubly excited states\cite{
rungeDensityFunctionalTheoryTimeDependent1984,
petersilkaExcitationEnergiesTimeDependent1996}.
To address this issue,
the spin-flip TD-DFT\cite{
wangTimedependentDensityFunctional2004,
minezawaOptimizingConicalIntersections2009,
huix-rotllantAssessmentNoncollinearSpinflip2010}
that describes double excitations as 
excitations from high-spin reference states was developed.
However,
choosing a suitable high-spin reference state is a nontrivial task 
that has high requirement of chemical intuition\cite{
huix-rotllantAssessmentNoncollinearSpinflip2010}.
To go beyond the adiabatic approximation,
a dressed TD-DFT method employing frequency-dependent 
exchange-correlation (XC) kernels has been proposed\cite{
maitraDoubleExcitationsTimedependent2004,
caveDressedTDDFTTreatment2004,
romanielloDoubleExcitationsFinite2009,
sangalliDoubleExcitationsCorrelated2011}.
The similar idea of adopting frequency-dependent XC kernels 
is also used
in the Bethe-Salpeter equation (BSE) formalism\cite{
salpeterRelativisticEquationBoundState1951,
blaseBetheSalpeterEquation2018,
blaseBetheSalpeterEquation2020}
to describe doubly excited states\cite{
sangalliDoubleExcitationsCorrelated2011,
zhangDynamicalSecondorderBetheSalpeter2013,
authierDynamicalKernelsOptical2020,
bintrimFullfrequencyDynamicalBethe2022}.
In addition to these approaches,
the ensemble DFT using weight-dependent local exchange functional\cite{
sagredo2018accurate,
marut2020weight} and orbital-optimized DFT\cite{
zhang2023target,haitExcitedStateOrbital2020} 
have also been used to compute double excitation energies.

Approaches based on the wave function theory (WFT) also face challenges
when describing doubly excited states.
%=====CI METHODS FOR SMALL SYSTEMS=====================================
For small systems,
highly accurate but computationally expensive methods such as 
the full configuration interaction (FCI),
Monte Carlo CI (MCCI)\cite{coeEfficientComputationTwoElectron2022},
semi-stochastic heat-bath CI (SHCI)\cite{holmesExcitedStatesUsing2017,
chienExcitedStatesMethylene2018},
iterative CI(iCI)\cite{zhangIterativeConfigurationInteraction2020a},
can be used to compute excitation energies with high accuracy.
%=====CASSCF AND VARIANTS==============================================
Multiconfigurational self-consistent field (MCSCF) methods 
and their perturbative variants,
including the complete-active-space self-consistent field 
(CASSCF),
complete-active-space perturbation theory (CASPT)\cite{
anderssonSecondorderPerturbationTheory1990,
anderssonSecondorderPerturbationTheory1992},
and n-electron valence state perturbation theory (NEVPT)\cite{
angeliIntroductionNelectronValence2001a,
angeliNelectronValenceState2001a,
angeliNelectronValenceState2002a,
angeliThirdorderMultireferencePerturbation2006},
treat all excited states equally,
and thus can naturally be used to calculate doubly excited states.
However, the computational cost grows exponentially with 
the number of active electrons and orbitals.
%=====NORMAL CC AND PT APPROACHES======================================
Coupled cluster methods (CC)\cite{cizekCorrelationProblemAtomic1966,
bartlettCoupledclusterTheoryQuantum2007,
loosMountaineeringStrategyExcited2018,
loosMountaineeringStrategyExcited2020}
and perturbation theory (PT) methods\cite{shavitt2009many}
are known for their ability to capture the dynamic correlation,
and thus are widely employed to calculate excited states.
To use these method for describing doubly excited states,
one has to go beyond the commonly used second-order approaches
so these methods can explicitly account for 
the two-particle-two-hole configurations.
Recently,
state-specific CC methods has been demonstrated to predict accurate
double excitation energies\cite{damourStateSpecificCoupledClusterMethods2024}.
%=====EOM-CC===========================================================
With the equation-of-motion (EOM) formalism,
the EOM-CC methods\cite{
christiansenSecondorderApproximateCoupled1995,
hattigCC2ExcitationEnergy2000,
purvisFullCoupledclusterSingles1982,
scuseriaClosedshellCoupledCluster1987,
kochCoupledClusterResponse1990,
stantonEquationMotionCoupledcluster1993}
provide a systematically improvable approach for predicting single and multi excitations.
Although EOM-CCSD includes the two-particle-two-hole configurations
in the calculations,
the accuracy for double excitations is not sufficient\cite{
loos2021accurate,loos2022mountaineering}.
Therefore,
one still needs to use higher-order methods within the EOM-CC scheme\cite{loos2021accurate,loos2022mountaineering,rishi2017excited},
which are computationally demanding.

%=pp-RPA===============================================================
In addition to the above approaches,
the particle-particle random-phase approximation (ppRPA) has 
the potential to serve as a more computationally efficient pathway 
to calculate double excitation energies.
%=====ppRPA CONSTRUCTION===============================================
The ppRPA,
which was originally developed to calculate the nuclear 
many-body correlation energy\cite{ring2004nuclear,blaizot1986quantum},
has been further developed to describe the electronic correlation energies, 
and electronic excitation energies
in molecular and bulk systems\cite{
vanaggelenExchangecorrelationEnergyPairing2013,
vanaggelenExchangecorrelationEnergyPairing2014,
yangDoubleRydbergCharge2013,
liAccurateExcitationEnergies2024,
liParticleParticleRandom2024}.
ppRPA can be equivalently derived from different perspectives 
such as the two-particle Green's function\cite{
blaizot1986quantum}
and the EOM ansatz\cite{ring2004nuclear,
yangBenchmarkTestsSpin2013}.
In the context of DFT,
ppRPA can be derived from the adiabatic connection\cite{
vanaggelenExchangecorrelationEnergyPairing2013,
vanaggelenExchangecorrelationEnergyPairing2014} 
as a counterpart of the commonly used 
particle-hole random phase approximation (phRPA)\cite{
bohmCollectiveDescriptionElectron1951,
renRandomphaseApproximationIts2012}
and from TD-DFT with the pairing field as the second 
order derivative of the XC energy with respect to 
the pairing matrix\cite{pengLinearresponseTimedependentDensityfunctional2014a},
which rationalize the practice of using Kohn-Sham orbitals 
and eigenvalues in ppRPA.
The ppRPA correlation energy is exact up to the second 
order in electron-electron interaction\cite{
vanaggelenExchangecorrelationEnergyPairing2013} 
and is equivalent to the ladder coupled-cluster doubles\cite{
pengEquivalenceParticleparticleRandom2013,
scuseriaParticleparticleQuasiparticleRandom2013}.
For calculations of excitation energies,
ppRPA can be considered as an approximation to double-electron-affinity 
or double-ionization-potential EOM-CC doubles\cite{
yangDoubleRydbergCharge2013,
berkelbachCommunicationRandomphaseApproximation2018}.
In ppRPA, the excitation energies of a given $N$-electron system 
can be calculated as the differences between the two-electron addition
energies of the $(N-2)$-electron system
from the particle-particle (pp) channel.
Similarly,
the excitation energies can also be obtained from the differences 
between the two-electron removal energies of 
the $(N+2)$-electron system from the hole-hole (hh) channel\cite{
yangDoubleRydbergCharge2013}.
The choice of the pp or the hh channels enhances 
the flexibility of the ppRPA method.

%=====ppRPA FEATURES AND PREVIOUS SUCCESS===============================
ppRPA has been further developed in many aspects and achieved a lot of success
in recent years due to its intrinsic features.
By the construction, 
ppRPA can be understood as a Fock-space embedding method 
that handles two electrons in a subspace CI manner, 
seamlessly integrated with DFT for the remaining $(N-2)$ electrons\cite{
yangSingletTripletEnergy2015,
zhangAccurateEfficientCalculation2016}. 
As a result, 
ppRPA offers a highly accurate description of the static correlation 
for the two non-bonding electrons in diradical systems,
and therefore predicts accurate singlet-triplet gaps 
in diatomic, carbene-like, disjoint and four-$\pi$-electron diradicals\cite{yangSingletTripletEnergy2015}.
In bulk systems, 
ppRPA has also been shown to accurately describe correlated 
excited states of point defects\cite{
liAccurateExcitationEnergies2024,
liParticleParticleRandom2024}.
Moreover, 
since the ppRPA kernel exhibits the correct long-range asymptotic behavior, 
it can accurately predict charge-transfer (CT)
and Rydberg excitation energies\cite{yangDoubleRydbergCharge2013,
yangChargeTransferExcitations2017}.
Recently, 
ppRPA has been employed in the multireference DFT approach, 
yielding accurate dissociation and excitation energies\cite{
chenMultireferenceDensityFunctional2017,
liMultireferenceDensityFunctional2022}.
The hole-hole channel in ppRPA has also been used to simulate
molecular dynamics properties\cite{yuInitioNonadiabaticMolecular2020,bannwarthHoleHoleTamm2020}.
The formal scaling of ppRPA for computing excitation energies is $\mathcal{O}(N^4)$ with the Davidson algorithm\cite{yangExcitationEnergiesParticleparticle2014}.
Development in using active space in ppRPA\cite{zhangAccurateQuasiparticleSpectra2017a} and particularly the recent truncation approach \cite{liLinearScalingCalculations2023}
has significantly reduced the computational cost without loss of accuracy,
making wider applications of the method possible.
In the Green's function theory, 
ppRPA eigenvalues and eigenvectors have also been utilized to construct the self-energy
in the T-matrix approximation to calculate 
quasiparticle energies\cite{zhangAccurateQuasiparticleSpectra2017a,
liRenormalizedSinglesGreens2021,
orlandoThreeChannelsManybody2023,
marie2024anomalous}.

%=====ppRPA AND DOUBLE EXCI============================================
Compared to particle-hole formalisms,
one of the major advantages of ppRPA is that it incorporates information from 
the particle-particle and hole-hole channels, 
which makes it naturally adept at describing double excitations.
It has been shown that ppRPA with traditional XC functionals 
predicts accurate double excitation energies for small molecules 
with errors around 0.5 eV\cite{
yangDoubleRydbergCharge2013,
zhangAccurateEfficientCalculation2016}, 
and also for polymer systems\cite{sutton2018single}.
Recently, ppRPA has been further applied to analyze doubly 
excited states in point defects\cite{
liParticleParticleRandom2024}.
In this work,
we explore the effectiveness and robustness of 
ppRPA for 
providing accurate double excitation energies.
We examine the performance of ppRPA based on a broad range of
XC functionals for describing double excitations with a comprehensive 
benchmark set developed in Refs~\citenum{
loosReferenceEnergiesDouble2019,kossoskiReferenceEnergiesDouble2024a},
which contains excited states with genuine and partial double 
excitation characters.
In a new direction, 
we also developed the ppRPA calculations 
from an excited state calculated within  the $\Delta$SCF framework. 
In particular, in the ppRPA@B3LYP calculation for acrolein,
the ($N-2$)-electron system is obtained by removing two electron from the
HOMO$-$1 orbital of the $N$-electron system.
This is as well justified as ppRPA calculations 
starting from on a DFT ground state, 
given the recent theoretical work establishing 
the foundation of the $\Delta$SCF method\cite{yang2024foundation}.
It should open up possibility of describing excited states 
that are not traditionally accessible.

%======================================================================
\section{THEORY}
%======================================================================
Similar to phRPA that is formulated with the density-density response function,
ppRPA is formulated with the particle-particle propagator
describing the dynamic fluctuation of the pairing matrix 
$\left \langle \Psi_0^N | \hat{a}_p \hat{a}_q | \Psi_0^N \right \rangle$,
where $\Psi_0^N$ is the wave function of the $N$-electron reference state
and $\hat{a}_p$ is the annihilation operator 
in the second-quantization notation.
In this paper,
we use 
$p$, $q$, $r$, $s$ for general molecular orbitals, 
$i$, $j$, $k$, $l$ for occupied orbitals, 
$a$, $b$, $c$, $d$ for virtual orbitals, 
and $m$ for the index of the two-electron addition/removal energy.
In the frequency space, 
the time-ordered pairing matrix fluctuation is\cite{vanaggelenExchangecorrelationEnergyPairing2013,vanaggelenExchangecorrelationEnergyPairing2014} 
\begin{equation}\label{eq:pairing_matrix}
    K_{pqrs}(\omega) = 
        \sum_m \frac{
            \left \langle \Psi^N_0 | \hat{a}_p \hat{a}_q | \Psi^{N+2}_m \right \rangle 
            \left \langle \Psi^{N+2}_m | \hat{a}_s^{\dagger} \hat{a}_r^{\dagger} | \Psi^N_0\right\rangle 
            }{
            \omega - \Omega^{N+2}_m + i\eta
            }
        - \sum_m \frac{
            \left \langle \Psi^N_0 | \hat{a}_s^{\dagger} \hat{a}_r^{\dagger} | \Psi^{N-2}_m\right\rangle 
            \left \langle \Psi^{N-2}_m | \hat{a}_p \hat{a}_q | \Psi^N_0  \right\rangle 
            }{
            \omega - \Omega^{N-2}_m - i\eta 
            }
\end{equation}
where $\hat{a}_p^{\dagger}$ is the second quantization creation operator, 
$\Omega^{N\pm2}$ is the two-electron addition/removal energy, 
and $\eta$ is a positive infinitesimal number.

The pairing matrix fluctuation $K$ of the interacting system 
can be approximated from the non-interacting $K_0$ with the Dyson equation\cite{vanaggelenExchangecorrelationEnergyPairing2013,vanaggelenExchangecorrelationEnergyPairing2014}
\begin{equation}\label{eq:dyson}
    K = K^0 + K^0 V K
\end{equation}
where the interaction 
$V_{pqrs}
    = \langle pq | rs \rangle - \langle pq | sr \rangle$ 
is used and 
$\langle pq|rs\rangle 
    = \int dx_{1}dx_{2}
        \frac{
        \phi_{p}^{*}(x_{1})\phi_{q}^{*}(x_{2})\phi_{r}(x_{1})\phi_{s}(x_{2})
        }{
        |r_{1}-r_{2}|
        }$.
In the direct ppRPA,
the exchange term in $V$ in Eq.~\ref{eq:dyson} is neglected\cite{tahirComparingParticleparticleParticlehole2019}.

Eq.~\ref{eq:dyson} can be written as a generalized eigenvalue equation\cite{vanaggelenExchangecorrelationEnergyPairing2013,vanaggelenExchangecorrelationEnergyPairing2014},
with a form similar to the Casida equation in TD-DFT
\begin{equation}\label{eq:eigen_equation}
\begin{bmatrix}\mathbf{A} & \mathbf{B}\\
\mathbf{B}^{\text{T}} & \mathbf{C}
\end{bmatrix}\begin{bmatrix}\mathbf{X}\\
\mathbf{Y}
\end{bmatrix}=\Omega^{N\pm2}\begin{bmatrix}\mathbf{I} & \mathbf{0}\\
\mathbf{0} & \mathbf{-I}
\end{bmatrix}\begin{bmatrix}\mathbf{X}\\
\mathbf{Y}
\end{bmatrix}
\end{equation}
with
\begin{align}
A_{ab,cd} & =\delta_{ac}\delta_{bd}(\epsilon_{a}+\epsilon_{b})+\langle ab||cd\rangle \\
B_{ab,kl} & =\langle ab||kl\rangle \\
C_{ij,kl} & =-\delta_{ik}\delta_{jl}(\epsilon_{i}+\epsilon_{j})+\langle ij||kl\rangle 
\end{align}
where $a<b$, $c<d$, $i<j$, and $k<l$.
In the ppRPA calculations,
the DFT SCF calculations of the ($N\pm2$)-electron states at 
the ground-state geometry of $N$-electron system are first performed,
then the orbitals and corresponding orbital energies
are used in Eq.~\ref{eq:eigen_equation} 
for calculating two-electron addition energies.
The excitation energy is obtained from 
the difference between the lowest 
and a higher two-electron addition/removal energy.

For closed-shell ($N-2$)-electron systems,
Eq.~\ref{eq:eigen_equation} can be expressed in the spin-adapted form\cite{yangBenchmarkTestsSpin2013}.
The singlet ppRPA matrix is given by 
\begin{align}
A^{\text{s}}_{ab,cd} & =\delta_{ac}\delta_{bd}(\epsilon_{a}+\epsilon_{b}) 
+ \frac{1}{\sqrt{(1+\delta_{ab})(1+\delta_{cd})}} 
(\langle ab|cd\rangle + \langle ab|dc\rangle) 
\label{eq:a_singlet}\\
B^{\text{s}}_{ab,kl} & = \frac{1}{\sqrt{(1+\delta_{ab})(1+\delta_{kl})}} 
(\langle ab|kl\rangle + \langle ab|lk\rangle) 
\label{eq:b_singlet}\\
C^{\text{s}}_{ij,kl} & =-\delta_{ik}\delta_{jl}(\epsilon_{i}+\epsilon_{j})
+ \frac{1}{\sqrt{(1+\delta_{ij})(1+\delta_{kl})}} 
(\langle ij|kl\rangle + \langle ij|lk\rangle) \label{eq:c_singlet}
\end{align}
with  $a \leq b$, $c \leq d$, $i \leq j$ and $k \leq l$.
The triplet ppRPA matrix is given by
\begin{align}
A^{\text{t}}_{ab,cd} & =\delta_{ac}\delta_{bd}(\epsilon_{a}+\epsilon_{b})
+\langle ab||cd\rangle 
\label{eq:a_triplet}\\
B^{\text{t}}_{ab,kl} & =\langle ab||kl\rangle 
\label{eq:b_triplet}\\
C^{\text{t}}_{ij,kl} & =-\delta_{ik}\delta_{jl}(\epsilon_{i}+\epsilon_{j})
+\langle ij||kl\rangle 
\label{eq:c_triplet} 
\end{align}

To reduce the computational cost,
an active space composed of $N_{\text{occ,act}}$ occupied and 
$N_{\text{vir,act}}$ virtual orbitals can be employed 
to construct ppRPA matrices\cite{liLinearScalingCalculations2023}.
As a result, 
the indices of the singlet ppRPA matrices 
in Eq.~\ref{eq:a_singlet} to Eq.~\ref{eq:c_singlet}
are constrained as
\begin{align}
    & a \leq b \leq N_{\text{vir,act}} \text{ and } 
        c \leq d \leq N_{\text{vir,act}}  \\
    & i \leq j \leq N_{\text{occ,act}} \text{ and } 
        k \leq l \leq N_{\text{occ,act}}
\end{align}
For the triplet ppRPA matrix in Eq.~\ref{eq:a_triplet} to Eq.~\ref{eq:c_triplet}, 
the indices are constrained in a similar way.
In this work,
the active-space ppRPA is used in calculations for point defects.
As shown in Ref.~\citenum{liLinearScalingCalculations2023},
the scaling of active-space ppRPA is $\mathcal{O} (N_\mathrm{act}^4)$ 
with the Davidson algorithm, 
where $N_\mathrm{act}$ is the number of orbitals in the active space. 
In addition, 
two-electron integrals can be efficiently constructed with 
the resolution of identity or the density-fitting (DF) technique.
For the three-index DF matrix, 
the atomic orbital (AO)-to-molecular orbital (MO) transformation step 
scales as $\mathcal{O} (N_\mathrm{aux} N^2_\mathrm{AO} N_\mathrm{act})$,
where $N_\mathrm{AO}$ is the number of atomic orbitals
and $N_\mathrm{aux}$ is the number of auxiliary basis functions.

%======================================================================
\section{COMPUTATIONAL DETAILS}
%======================================================================
For double excitation energies of molecular systems,
the geometries are from Ref.~\citenum{kossoskiReferenceEnergiesDouble2024a}.
The SCF calculations of the ($N-2$)-electron systems were performed
with the aug-cc-pVTZ basis set\cite{kendall1992electron,woon1993gaussian}
using the PySCF package\cite{sunPySCFPythonbasedSimulations2018,
sunRecentDevelopmentsPySCF2020}.
All the calculations were performed in the full space with 
the Davidson algorithm\cite{yangExcitationEnergiesParticleparticle2014}.
Results from aug-cc-pVQZ calculations are documented in 
the supporting information (SI).
Geometries of point defects were taken from Ref.~\cite{liParticleParticleRandom2024}.
For the nitrogen-vacancy (NV$^-$) in diamond,
the ($N+2$)-electron ground state was used and excitation energies 
were calculated within the hole-hole channel in ppRPA.
For the carbon vacancy (VC) in diamond,
the geometry with D$_{\text{2d}}$ symmetry was used, and
the ($N-2$)-electron ground state was used and excitation energies 
were calculated within the particle-particle channel in ppRPA.
All periodic defect calculations were performed with $\Gamma$-point sampling and Gaussian density fitting, with cc-pVDZ basis set~\cite{
dunningGaussianBasisSets1989} 
and the corresponding cc-pVDZ-RI auxiliary basis set~\cite{
weigendEfficientUseCorrelation2002}.
In this work,
all ground-state DFT calculations were carried out using 
the PySCF quantum chemistry software package~\cite{
sunPySCFPythonbasedSimulations2018,sunRecentDevelopmentsPySCF2020}. 

%======================================================================
\section{RESULTS}
%======================================================================
\subsection{\label{sec:mol}Double excitation of molecular systems}

%=Present results======================================================
Double excitation energies of 21 molecules from Ref.~\citenum{
kossoskiReferenceEnergiesDouble2024a} were
calculated with ppRPA.
Eight functionals,
including three hybrid functionals,
one generalized gradient approximation (GGA),
one meta-GGA,
and three range-separated hybrid functionals,
were used to perform SCF calculations
on the ($N-2$)-electron systems as starting points of 
ppRPA calculations.
Excitation energies, 
mean signed errors (MSEs), 
and mean absolute errors (MAEs) from these calculations
are tabulated in Table~\ref{tab:mol_double_e}.

%=Multi-ref feature of ppRPA===========================================
Two types of double excitations are investigated in this study: 
genuine double excitations (gd), 
which are primarily characterized by doubly excited configurations, 
and partial double excitations (pd), 
which exhibit significant contributions from singly excited configurations. 
This classification is based on CC3 calculations\cite{
kossoskiReferenceEnergiesDouble2024a}. 
ppRPA provides comparable accuracy in describing both types of double excitations.
This capability arises from ppRPA’s strength in calculating states 
with multi-reference character. 
Within ppRPA, 
two electrons are treated accurately in a CI framework, 
while the remaining 
$N-2$ electrons are described using DFT.

%=Comparison among functionals=========================================
Among the eight functionals we tested,
TPSSh produces the lowest MAE (0.350 eV) 
and $\omega$B97X-D produces the highest MAE (0.530 eV),
indicating a small starting point dependence in ppRPA for predicting
double excitation energies.
The difference between the two values is only 0.180 eV,
much smaller than the functional differences between DFT-based approaches reported 
in Refs~\citenum{marut2020weight,haitExcitedStateOrbital2020,zhang2023target}.
%=Accuracy and exact exchange==========================================
One of the reasons for the different accuracy of different functionals
may be the percentage of the exact exchange.
In Table~\ref{tab:exact_x},
we list the percentages of the exact exchange of the functionals studied.
%=====GGA and meta-GGA=================================================
For ppRPA@PBE and ppRPA@SCAN, 
where the functionals have no exact exchange, 
the MAEs are both around 0.40 eV.
The difference is that ppRPA@PBE underestimates 
the double excitation energies,
while ppRPA@SCAN overestimates them.
%=====hybrid functionals===============================================
For ppRPA calculations starting from SCF results obtained from hybrid
functionals,
ppRPA@M06-2X where the functional contains 54\% exact exchange 
produces the largest MAE and MSE.
With the functionals containing 20\% and 10\% exact exchange respectively,
ppRPA@B3LYP and ppRPA@TPSSh provides similar accuracy. 
The MAEs differ by around 0.01 eV,
and the MSEs differ by less than 0.1 eV.
%=====range-separated functionals======================================
Among the three range-separated hybrid functionals,
two of them,
namely CAM-B3LYP and $\omega$B97X-D,
have significant amount (65\% and 100\%) of exact exchange in the long range.
Both the MAEs from ppRPA@CAM-B3LYP and ppRPA@$\omega$B97X-D 
are over 0.5 eV,
which is similar to that from ppRPA@M06-2X.
For ppRPA@HSE03,
where the functional does not contain the exact exchange
in the long range,
the MAE is similar to those from ppRPA@B3LYP and ppRPA@TPSSh.
To conclude,
ppRPA starting from hybrid functionals with 
around 10\%$\sim$20\% exact exchange can provide more accurate 
double excitation energies compared to GGAs and meta-GGAs,
while increasing the amount of exact exchange may cause the 
accuracy to decrease.
This agrees with previous results of ppRPA for predicting 
double excitation energies of small molecules\cite{yangDoubleRydbergCharge2013}.
For range-separated functionals,
a high fraction of exact exchange in the long range
can lead to less accurate descriptions 
of double excitation energies.

%=Compare ppRPA with WFT methods in Loos's paper=======================
Here we compare the results form ppRPA with those from other methods.
In Ref.~\citenum{kossoskiReferenceEnergiesDouble2024a},
Kossoski et al. provide reference energies for double excitations 
from 11 WFT methods.
Comparing the MAEs from ppRPA and the 11 WFT methods,
we find that ppRPA starting from DFAs with proper amount of 
exact exchange is more accurate than 
state-averaged CASSCF (SA-CASSCF) and CC3,
whose MAEs are 0.48 eV and 0.56 eV respectively.
With carefully selected DFAs,
ppRPA provides similar accuracy as CCSDT and CASPT2
for double excitation energies.
Because of the computationally favorable cost,
ppRPA can be an efficient alternative to WFT methods for predicting 
double excitations.

\begin{figure}
\includegraphics[width=0.6\textwidth]{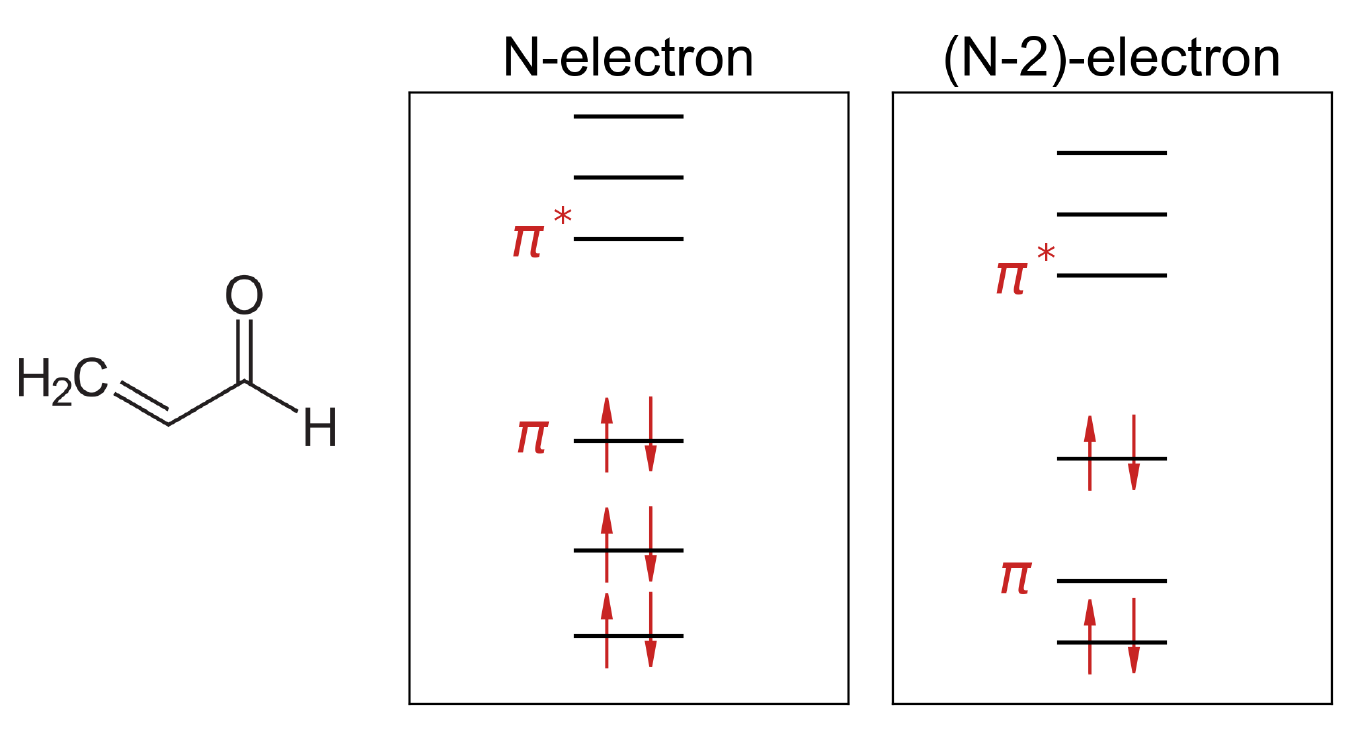}
\caption{Valence energy levels in $N$-electron and 
($N-2$)-electron systems of acrolein obtained from B3LYP 
(energy levels are qualitative only).
In the ($N-2$)-electron system,
two electrons are removed from the $\pi$ orbital.}
\label{fig:acrolein}
\end{figure}

%=Special case: acrolein from B3LYP calculation========================
In all the calculations performed in this work,
the $^1A'$ ($\pi \rightarrow \pi^*$) state of acrolein from 
ppRPA@B3LYP is a special one.
As shown in Fig.~\ref{fig:acrolein},
after taking two electrons away,
orbital misalignment happens in the ($N-2$)-electron system.
From the SCF calculation on the $N$-electron system,
one can know that the $\pi$ orbital from which the electrons 
are excited from is 
the highest occupied molecular orbital (HOMO).
Based on this,
one would intuitively take two electrons away from the HOMO
to form the ($N-2$)-electron system.
However,
in the ($N-2$)-electron system, 
the orders of the $\pi$ orbital and the orbital below it
are switched.
Therefore,
we used the $\Delta$SCF approach using the maximum overlap method (MOM)\cite{gilbert2008self}
implemented in PySCF to obtain the proper starting-point configuration
from $\Delta$SCF\cite{yang2024foundation}, 
which is as a doubly excited state of the ($N-2$)-electron system. 
The ppRPA equations for such calculations based on 
$\Delta$SCF approach remain the same as in Eqs. (3-12).

{\LTcapwidth=\textwidth \setlength\tabcolsep{1pt} \fontsize{3}{4}\selectfont{
\begin{longtable}{@{\extracolsep{\fill}}llllcccccccc}
\caption{Vertical double excitation energies calculated with ppRPA 
using the aug-cc-pVTZ basis set.
TBE stands for theoretical best estimate.
MAE and MSE stand for mean absolute error and mean signed error.
pd and gd stand for partial double excitation and 
genuine double excitation \cite{kossoskiReferenceEnergiesDouble2024a}.
MAEs and MSEs are calculated with respect to TBEs.
Some results are not available (denoted as N/A) due to the convergence 
issue of the ($N-2$)-electron systems.
All values are in eV.}
\label{tab:mol_double_e}\\
\toprule
                     &         &    &       & \multicolumn{8}{c}{functional} \\
\cline{4-11} 
system               & state   &type& TBE   & B3LYP  & M06-2X & TPSSh  & PBE    & SCAN   & CAM-B3LYP & HSE03  & $\omega$B97X-D \\
\midrule
\endfirsthead
\multicolumn{10}{l}{TABLE \ref{tab:mol_double_e} continued:}\\
\toprule
system               & state   &type& TBE   & B3LYP  & M06-2X & TPSSh  & PBE    & SCAN   & CAM-B3LYP & HSE03  & $\omega$B97X-D \\
\midrule
\endhead
\bottomrule
\endfoot
\bottomrule
\endlastfoot
acrolein             & $^1A'$  &pd& 7.928 & 7.945  & N/A    & N/A    & N/A    & N/A    & 8.159     & N/A    & 8.183   \\
benzoquinone         & $^1A_g$ &gd& 4.566 & 5.976  & 6.753  & 5.565  & 4.904  & 5.461  & N/A       & 6.098  & N/A     \\
borole               & $^1A_1$ &pd& 6.484 & 7.395  & 7.388  & 7.481  & N/A    & 7.640  & 7.469     & 7.566  & 7.520   \\
                     & $^1A_1$ &gd& 4.708 & 4.656  & 4.809  & 4.702  & N/A    & 4.757  & 4.787     & 4.765  & 4.858   \\
butadiene            & $^1A_g$ &pd& 6.515 & 6.484  & 6.612  & 6.506  & 6.162  & 6.595  & 6.701     & 6.630  & 6.748   \\
cyclobutadiene       & $^1A_g$ &gd& 4.036 & 4.018  & 4.068  & 4.084  & 3.923  & 4.158  & 4.058     & 4.096  & 4.114   \\
cyclopentadiene      & $^1A_1$ &pd& 6.451 & 6.489  & 6.653  & 6.534  & 6.238  & 6.663  & 6.652     & 6.627  & 6.673   \\
cyclopentadienethione& $^1A_1$ &pd& 5.329 & N/A    & 4.269  & N/A    & N/A    & N/A    & 4.310     & N/A    & 4.206   \\
cyclopentadienone    & $^1A_1$ &pd& 6.714 & 7.691  & 7.686  & N/A    & N/A    & N/A    & 7.762     & N/A    & 7.809   \\
                     & $^1A_1$ &gd& 5.009 & 5.363  & 5.533  & N/A    & N/A    & N/A    & 5.488     & N/A    & 5.439   \\
diazete              & $^1A_1$ &gd& 6.605 & 6.733  & N/A    & N/A    & N/A    & N/A    & 6.755     & 6.872  & 6.886   \\
ethylene             & $^1A_g$ &gd& 12.899& 12.737 & 12.204 & 12.985 & 12.968 & 13.491 & 12.494    & 12.880 & 12.641  \\
formaldehyde         & $^1A_1$ &gd& 10.426& 10.371 & 10.089 & 10.672 & 10.398 & 10.832 & 10.245    & 10.498 & 10.479  \\
glyoxal              & $^1A_g$ &gd& 5.492 & 5.810  & 6.329  & 5.584  & 4.982  & 5.628  & 6.374     & 5.947  & 6.247   \\
hexatriene           & $^1A_g$ &pd& 5.435 & 5.046  & 5.581  & 4.964  & 4.546  & 4.921  & 5.560     & 5.169  & 5.603   \\
naphthalene          & $^1A_g$ &pd& 6.748 & 6.414  & 7.067  & 6.326  & 7.408  & 6.283  & 7.000     & 6.573  & 6.997   \\
nitrosomethan        & $^1A'$  &gd& 4.732 & 4.247  & 4.147  & 4.296  & 4.412  & 4.455  & 4.202     & 4.299  & 4.239   \\
nitrous acid         & $^1A'$  &gd& 7.969 & 8.528  & 8.647  & 8.572  & 8.247  & 8.690  & 8.627     & 8.629  & 8.666   \\
nitroxyl             & $^1A'$  &gd& 4.333 & 4.638  & 4.624  & 4.708  & 4.575  & 4.757  & 4.639     & 4.699  & 4.732   \\
octatetraene         & $^1A_g$ &pd& 4.68  & 4.140  & 4.770  & 4.013  & 3.594  & 3.907  & 4.744     & 4.235  & 4.809   \\
oxalyl fluoride      & $^1A_1$ &gd& 8.923 & 9.556  & 9.299  & 9.373  & 8.937  & 9.619  & 9.500     & 9.786  & 9.771   \\
pyrazine             & $^1A_g$ &pd& 8.48  & 8.679  & 9.000  & N/A    & N/A    & N/A    & 9.060     & 8.741  & 9.025   \\
                     & $^1A_g$ &gd& 7.904 & 8.607  & N/A    & 8.430  & N/A    & N/A    & 8.921     & N/A    & 8.914   \\
tetrazine            & $^1A_g$ &gd& 4.951 & 5.216  & 5.751  & 4.984  & 4.467  & 4.875  & 5.693     & 5.333  & 5.621   \\
                     & $^1B_3$ &gd& 6.215 & 7.004  & 7.530  & 6.791  & 6.152  & 6.756  & 7.514     & 7.155  & 7.429   \\
                     & $^3B_3$ &gd& 5.848 & 5.989  & 6.577  & 5.753  & 5.083  & 5.702  & 6.536     & 6.145  & 6.437   \\
\midrule
         \multicolumn{4}{c}{MSE}       & 0.227  & 0.367  & 0.150  & -0.190 & 0.199  & 0.342     & 0.297  & 0.374   \\
         \multicolumn{4}{c}{MAE}       & 0.393  & 0.600  & 0.361  & 0.378  & 0.436  & 0.520     & 0.425  & 0.530  
\end{longtable}
}}
\FloatBarrier

\begin{table}[]
\caption{Percentages of the exact exchange in various functionals.}
\label{tab:exact_x}
\setlength\tabcolsep{20pt}
\begin{tabular}{lc}
\toprule
functional        & exact exchange\%                     \\
\midrule
B3LYP             & 20\%                                 \\
M06-2X            & 54\%                                 \\
TPSSh             & 10\%                                 \\
PBE               & 0\%                                  \\
SCAN              & 0\%                                  \\
CAM-B3LYP         & short range 19\%, long range 65\%    \\
HSE03             & short range 25\%, long range 0\%     \\
$\omega$B97X-D    & short range 22.2\%, long range 100\% \\
\bottomrule
\end{tabular}
\end{table}
\FloatBarrier

\begin{figure}
\includegraphics[width=0.6\textwidth]{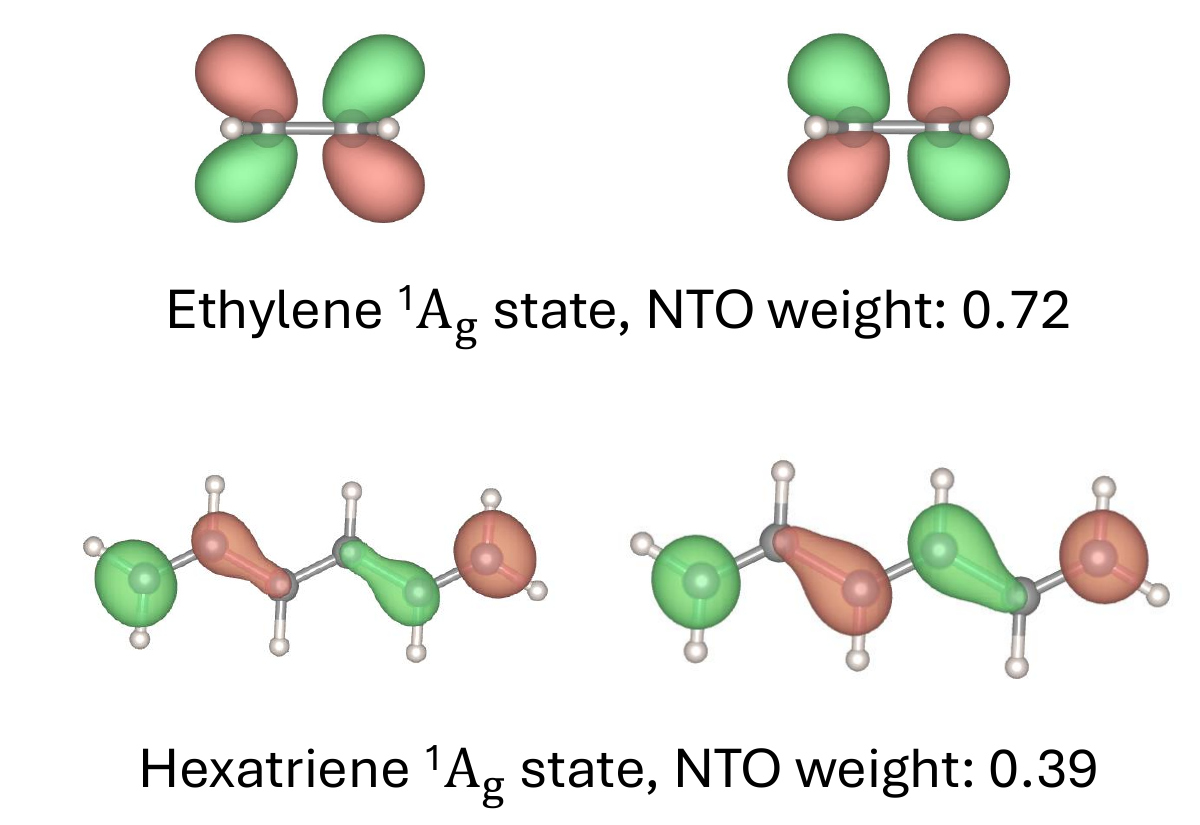}
\caption{Natural transition orbitals of double excitations for the $^1$A$_\text{g}$ state 
in ethylene and the $^1$A$_\text{g}$ state in hexatriene obtained from ppRPA@M06-2X.
The aug-cc-pVTZ basis set was used.
Isosurface value is 0.06 a.u.}
\label{fig:nto}
\end{figure}

We further visualize double excitations obtained from ppRPA 
using natural transition orbitals (NTOs) developed in Ref.\citenum{liParticleParticleRandom2024}.
NTOs of double excitations for the $^1$A$_\text{g}$ state 
in ethylene and the $^1$A$_\text{g}$ state in hexatriene are shown in Fig.~\ref{fig:nto}.
For the $^1$A$_\text{g}$ state in ethylene,
NTOs corresponding to ($\pi^*,\pi^*$) has a NTO weight of 0.72,
which agrees with its genuine double excitation nature.
For the partial double excited $^1$A$_\text{g}$ state in hexatriene,
NTOs corresponding to ($\pi^*,\pi^*$) provides a smaller NTO weight of 0.39.
\FloatBarrier

%======================================================================
\subsection{Double excitations in point defects}
\FloatBarrier

\begin{figure}
\includegraphics[width=0.6\textwidth]{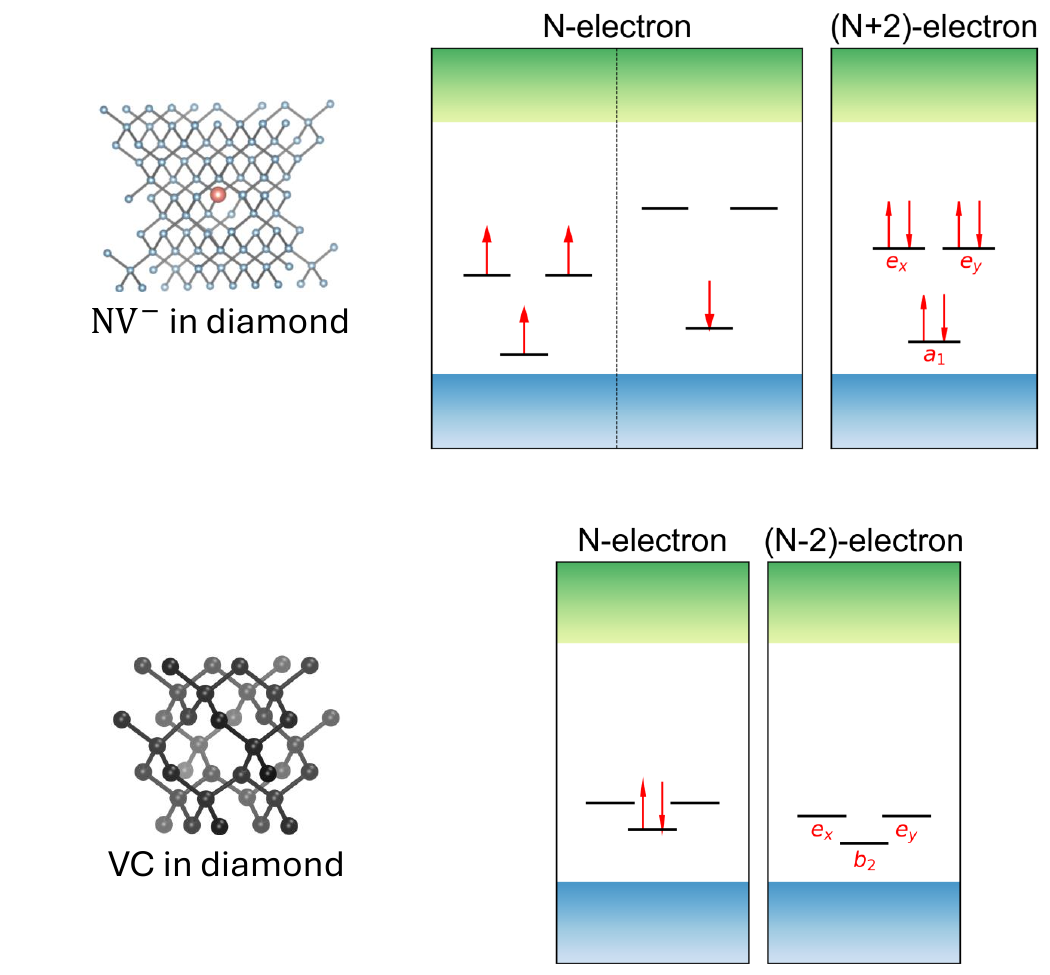}
\caption{Illustration of defect energy levels and ground-state electron configurations of NV$^-$ and VC in diamond (energy levels are qualitative only).}
\label{fig:defect_level}
\end{figure}

\begin{table}
\setlength\tabcolsep{15pt}
\caption{Vertical excitation energies and dominant configuration contributions of defect states involving with double excitation characters in NV$^-$ and VC in diamond obtained from the ppRPA approach based on different functionals.
The geometry with D$_{\text{2d}}$ symmetry was employed for VC in diamond.
The cc-pVDZ basis set was used. 
All values of excitation energies are in \,{eV}.}\label{tab:defect}
\begin{tabular}{ccccc}
\toprule
NV$^-$ in diamond & $^1$A$_1$ excitation & \multicolumn{3}{c}{dominant configuration in $^1$A$_1$} \\
\cmidrule(l{0.5em}r{0.5em}){2-2} \cmidrule(l{0.5em}r{0.5em}){3-5}
experiment~\cite{daviesOpticalStudies9451997,haldarLocalExcitationsCharged2023} & 1.76$\sim$1.85 & $|a_1 \bar{a}_1 e_x \bar{e}_x\rangle$ & $|a_1 \bar{a}_1 e_y \bar{e}_y\rangle$ & $|e_x \bar{e}_x e_y \bar{e}_y\rangle$ \\
PBE            & 1.67      & 33.4\%             & 33.4\%             & 11.0\%     \\
B3LYP          & 1.97      & 34.6\%             & 34.6\%             & 13.4\%     \\
CAM-B3LYP      & 2.09      & 33.8\%             & 33.8\%             & 16.5\%     \\
HSE03          & 1.96      & 34.9\%             & 34.8\%             & 12.0\%     \\
M06            & 2.08      & 33.5\%             & 33.5\%             & 15.0\%     \\
M06-2X         & 1.93      & 31.8\%             & 31.8\%             & 13.9\%     \\
SCAN           & 1.77      & 36.1\%             & 36.0\%             & 8.7\%      \\
TPSSh          & 1.83      & 35.0\%             & 34.9\%             & 10.1\%     \\
$\omega$B97X-D & 2.05      & 33.5\%             & 33.5\%             & 11.9\%     \\
\midrule
VC in diamond & $^1$E excitation & \multicolumn{3}{c}{dominant configuration in $^1$A$_1$} \\
\cmidrule(l{0.5em}r{0.5em}){2-2} \cmidrule(l{0.5em}r{0.5em}){3-5}
experiment~\cite{lannooOpticalAbsorptionNeutral1968}      & 2.20      & $|b_2 \bar{b}_2\rangle$ & $|e_x \bar{e}_x\rangle$ & $|e_y \bar{e}_y\rangle$ \\
PBE            & 1.75      & 90.0\%             & 6.0\%              & 6.0\%      \\
B3LYP          & 2.12      & 85.3\%             & 7.5\%              & 7.5\%      \\
CAM-B3LYP      & 2.35      & 83.3\%             & 8.5\%              & 8.5\%      \\
HSE03          & 2.09      & 85.4\%             & 7.5\%              & 7.5\%      \\
M06            & 2.16      & 83.3\%             & 8.4\%              & 8.4\%      \\
M06-2X         & 2.31      & 83.1\%             & 8.4\%              & 8.4\%      \\
SCAN           & 1.82      & 90.0\%             & 5.7\%              & 5.7\%      \\
TPSSh          & 1.87      & 88.2\%             & 6.4\%              & 6.4\%      \\
$\omega$B97X-D & 2.33      & 82.9\%             & 8.4\%              & 8.4\%      \\
\bottomrule
\end{tabular}
\end{table}

We further examine the performance of ppRPA for describing double excitations in periodic bulk systems.
Two point defect systems are tested: 
NV$^-$ in diamond and VC in diamond with D$_{2\text{d}}$ symmetry,
whose defect energy levels are shown in Fig.~\ref{fig:defect_level}.
Vertical double excitation energies of defect states involving with double excitation characters obtained from ppRPA based on different functionals compared with experiment values are tabulated in Table~\ref{tab:defect}.
To correct the finite supercell-size error, 
a linear fitting two-point supercell-size extrapolation scheme $E(1/N_\mathrm{atom}) = E_\infty + a/N_\mathrm{atom}$ was employed, 
which has been utilized to compute excitation energies of defect systems in the thermodynamic limit~\cite{jinExcitedStateProperties2023,vermaOpticalPropertiesNeutral2023,liAccurateExcitationEnergies2024,liParticleParticleRandom2024}.
For the NV$^-$ in diamond,
the hole-hole channel in ppRPA is used to calculate excitation energies.
The $^1$A$_1$ state in the NV$^-$ in diamond is shown to have a strong multireference character~\cite{liParticleParticleRandom2024,jinVibrationallyResolvedOptical2022}.
As shown in Table~\ref{tab:defect},
ppRPA based on different functionals provides similar descriptions for the $^1$A$_1$ state,
showing that the contribution of the doubly-excited configuration $|e_x \bar{e}_x e_y \bar{e}_y\rangle$ is around 15\%.
Excitation energies obtained from ppRPA with different functionals are around 1.8 to 2.1 \,{eV},
which shows a much smaller starting point dependence than TD-DFT\cite{liParticleParticleRandom2024,jinExcitedStateProperties2023}.
For the VC in diamond,
as shown in Ref.~\citenum{liParticleParticleRandom2024},
the ground state $^1$A$_1$ has a strong double excitation energy.
Single-reference methods fail to capture doubly-excited configurations $|e_x \bar{e}_x\rangle$ and $|e_y \bar{e}_y\rangle$,
which is the main source of large errors in TD-DFT\cite{liParticleParticleRandom2024}.
ppRPA based on different functionals shows that the $^1$A$_1$ state has 85\% contributions from the $|b_2 \bar{b}_2\rangle$ configuration as well as 15\% contributions from doubly-excited configurations $|e_x \bar{e}_x\rangle$ and $|e_y \bar{e}_y\rangle$.
With a proper description for the ground state,
ppRPA based on local and non-local functionals provides errors around 0.5 \,{eV} and 0.1 \,{eV} for the excitation energy of the $^1$E state, 
which are much smaller than the 1 \,{eV} error in TD-DFT\cite{liParticleParticleRandom2024}.
For defect systems,
ppRPA based on B3LYP and HSE03 provide smallest errors for excitation energies involving with double excitation characters,
agreeing well with the results for molecular systems in Section~\ref{sec:mol}.

\FloatBarrier

\section{CONCLUSIONS}
In this study, 
we demonstrated ppRPA as 
an accurate and computationally efficient method 
for calculating double excitation energies in molecular and bulk systems. 
Our benchmark of 21 molecular systems using different DFAs revealed that 
ppRPA with functionals containing 10-20\% exact exchange 
offers the best accuracy for molecular systems. 
Errors of ppRPA based on B3LYP, TPSSh and HSE03 are only around 0.40 eV,
comparable to computationally demanding WFT methods.
In particular,
in the calculation of ppRPA@B3LYP for acrolein,
the doubly excited state was obtained from ppRPA starting with a non-ground state determinant,
which opens up new possibilities for calculating excitation energies.
Additionally, 
ppRPA was shown to be effective in 
describing correlated states with double excitation characters in point defects on the equal footing.
This work establishes ppRPA as an accurate and low-cost tool for investigating double excitations of both molecular and periodic systems.

\section*{SUPPORTING INFORMATION}
See the Supporting Information for 
excitation energies of point defect systems 
and double excitation energies of molecular systems using
aug-cc-pVQZ basis set
obtained from ppRPA.

\begin{acknowledgments}
J.Y. and W.Y. acknowledge the support from the National Science Foundation (Grant No.~CHE-2154831).
T.Z. and J.L. are supported by the National Science Foundation (Grant No.~CHE-2337991) and a start-up fund from Yale University. 
J.L. also acknowledges support from the Tony Massini Postdoctoral Fellowship in Data Science. 
\end{acknowledgments}

\section*{Data Availability Statement}
The data that support the findings of this study are available from the corresponding author
upon reasonable request.

\bibliography{ref,extra,DoubleExci,ppRPA}

\end{document}